\documentclass[12pt]{article}
\usepackage{amsmath}
\usepackage{amssymb}
\usepackage{graphicx}

\numberwithin{equation}{section}

\newcommand{\Ncal}{\mathcal{N}}

\DeclareMathOperator*{\Tr}{{\rm Tr}}

\newcommand{\Fcal}{{\cal F}}
\newcommand{\Wcal}{{\cal W}}
\newcommand{\Mcal}{{\cal M}}

\begin{document}
\begin{titlepage}

 \renewcommand{\thefootnote}{\fnsymbol{footnote}}
\begin{flushright}
 \begin{tabular}{l}
 TU-819 \\
 RIKEN-TH-133\\
 \end{tabular}
\end{flushright}

 \vfill
 \begin{center}
{\Large Extended MQCD and SUSY/non-SUSY duality}

\vspace{1.9cm}

\noindent{{\normalsize Kazutoshi Ohta$^a$\footnote{E-mail:kohta@phys.tohoku.ac.jp} 
  and Ta-Sheng Tai$^b$\footnote{E-mail:tasheng@riken.jp} }}
\bigskip
\vspace{.1cm}
\centerline{\it $^a$ Department of Physics, Tohoku University, Sendai 980-8578, JAPAN}

\bigskip
 \centerline{\it $^b$ Theoretical Physics Laboratory, RIKEN,
Wako, Saitama 351-0198, JAPAN}
\vspace{.4cm}
\end{center}
 \vfill
\vspace{0.7cm}

\begin{abstract}
We study the SUSY/non-SUSY duality proposed by 
Aganagic et al. from 
Type IIA string and M-theory perspectives. We find that 
our brane configuration generalizes the so-called 
$extended$ Seiberg-Witten theory on the one hand, and 
provides a way to realize non-SUSY vacua by intersecting NS5-branes on the other hand. 
We also argue how the partial SUSY breaking from $\Ncal=2$ down to $\Ncal=1$ 
can be clearly visualized through the brane picture. 

\end{abstract}

\vfill
\vspace{0.5 cm}

\setcounter{footnote}{0}
\renewcommand{\thefootnote}{\arabic{footnote}}
\end{titlepage}

\newpage


\section{Introduction}
Recently, Aganagic et al. proposed a SUSY/non-SUSY duality \cite{Vafa 2008} 
in Type IIB string compactification. 
In contrast to 
previous works \cite{Vafa 2006, Mar,Vafa 2007} where anti-branes are introduced by hand, 
the breakthrough is to turn on a holomorphic varying background NS-flux $H_0$ through the non-compact Calabi-Yau (CY) three-fold. 
This soon suggests a way to realize 
various kinds of SUSY or non-SUSY vacua 
via adjusting parameters the NS-flux contains
\footnote{Applications and generalizations of these flux vacua are also discussed in a recent paper \cite{Hollands:2008cs}.}.

Let us briefly review their ideas. Because of the flux $H_0=dB_0$, 
four-dimensional gauge theory, realized by 
wrapping D5-branes on vanishing two-cycles of a CY, acquires 
different gauge couplings at each $\mathbb{P}^1$ locus: 
\begin{align}
\begin{aligned}
\alpha= \frac{\theta}{2\pi}+ 
\frac{4\pi i}{g^2_{YM}}= \int_{\mathbb{P}^1} B_0 (v) , ~~~~~
B_0 = B_{RR} + \frac{i}{g_s}B_{NS},
\label{B}
\end{aligned}
\end{align}
where 
$v$ parameterizes a CY bearing, say, the $A_1$-type singularity as 
\begin{align}
\begin{aligned}
X: uz + w^2-W'(v)^2=0.
\label{X}
\end{aligned}
\end{align}
Note that $W(v)=\sum^{n+1}_{k= 1}a_k v^{k}$, providing a non-trivial $A_1$ 
fibration over $v$, 
corresponds to the tree-level superpotential 
breaking $\Ncal=2$ down to $\Ncal=1$. Also, the adjoint chiral field $\Phi$ 
on D5-branes gets identified with the transverse $v$-direction. 
Although generalizing $X$ to other ALE fibrations 
can be carried out, 
the above prototype will prove to be sufficiently good 
due to arbitrarily many degrees of freedom inside $B_0 (v)$.

The proposed SUSY/non-SUSY duality is achieved by 
tuning coefficients of 
the $v$-dependent background $B$-field, which has the following expression%
\footnote{As noted in \cite{Vafa 2008}, the degree of $B_0 (v)$ polynomial is restricted to 
at most $n-1$ for triviality of the operator 
$\Tr \big( \Phi^k W'(\Phi) \Wcal^\alpha \Wcal_\alpha \big)$ in $\Ncal=1$ 
gauge theory chiral ring. }
\begin{align}
\begin{aligned}
\Fcal''_{UV}(v)=B_0 (v)=\sum^{n-1}_{k= 0} t_k v^k,
\label{F}
\end{aligned}
\end{align}
where $\Fcal_{UV}(v)$ denotes the ultraviolet prepotential%
\footnote{In $\Ncal=2$ gauge theory, the bare coupling constant $\alpha(\Phi)$ 
is determined by a holomorphic function $\Fcal_{UV}$ as $\alpha(\Phi)=\Fcal''_{UV}(\Phi)$. }. 
For generic $t_k$, SUSY is spontaneously broken at UV. 
This is accounted for by \eqref{B}, in which one observes that 
$\mathbb{P}^1$'s may develop relatively different 
orientations at critical points 
$W'(v)=\prod^{n}_{i=1} (v-v_i)=0$ for 
$\int B_{NS}\sim$ K\"ahler moduli of $\mathbb{P}^1$. On the other hand, some 
specific choice of $t_k$ can still make 
four supercharges preserved, i.e. all orientations of $\mathbb{P}^1$'s are kept aligned. 
As shown in \cite{Vafa 2008}, 
through geometric transition to dual CY manifolds, SUSY breaking 
effects can as well be captured qualitatively by studying 
strongly-coupled IR physics. Minimizing the effective 
superpotential there, one can further determine 
$t_k $ from $a_k $%
\footnote{This fact can be interpreted 
from the M-theory perspective, see below. }.

Like the brane realization \cite{Ooguri:2006bg,me1,me2} of meta-stable 
SUSY breaking vacua \cite{Intriligator:2006dd},
our purpose in this paper is to 
translate things considered above into Type IIA/M-theory language. 
It is well-known that via a T-duality acting on $X$ one instead obtains two NS5-branes 
in flat spacetime with 
D4-branes in between them. 
From the tree-level F-term 
\begin{align}
\begin{aligned}
 \int d^2\theta ~ \Fcal_{UV}'' (\Phi)\Wcal_\alpha \Wcal^\alpha + W(\Phi),
\label{Fterm}
\end{aligned}
\end{align}
one can choose a vacuum $\Phi={\it diag} (v_1, \cdots,v_2,\cdots, \cdots, v_n, \cdots)$ 
such that the gauge group $U(N)$ is broken to $\prod_{i=1}^n U(N_i)$. 
Then, it is seen that D4-branes, coming from fractional D3-branes, remain at $v_i$'s. 
The size and orientation of $\mathbb{P}_i^1$ controlled by \eqref{F} are translated, 
respectively, to 
the length along the T-dual direction (bare gauge coupling) and sign of RR charge of 
$i$-th stack of D4-branes. 
Based on this Type IIA tree-level 
description%
\footnote{As usual, we notice that tree-level 
field theory results match with 
classical brane pictures 
at the lowest order in $\ell_s$
under $g_s\to 0$, i.e. brane bending and string interaction are not taken into account.},  
$B_{NS}(v)<0$ which naively means negative gauge couplings can be understood as 
two crossing NS5-branes that result in anti-branes. How spontaneously SUSY breaking vacua occur 
can therefore be visualized clearly in the presence of both 
D4- and $\overline{\text{D4}}$-branes as a consequence of the 
$extended$ prepotential.

The rest of this paper is organized as follows. 
In the next section, we review some known facts about Type IIA/M-theory 
brane configurations. 
In section 3, we study the SUSY/non-SUSY duality by introducing a varying $B$-field. 
We also comment on the partial SUSY breaking mechanism 
in terms of Type IIA brane pictures. 
Finally, we conclude in Section 4.

\section{Type IIA/M-theory brane picture} 

To set up notations in this paper, we briefly review 
Type IIA/M-theory brane configurations here%
\footnote{For more details, see \cite{GK} and references therein.}. 

\subsection{Type IIA setup}

Viewing alternatively $X$ in \eqref{X} as an $U(1)$ fiber over $(v,w)$-plane, one can go from Type IIB CY geometry 
to Type IIA Hanany-Witten \cite{HW} type brane setup  
upon a T-daulity along this $S^1$ ($x^6$-direction) \cite{a,a1,a2,a3}, namely, 
\begin{align}
\begin{aligned}
(u,z,w,v)\to(\lambda u,\lambda^{-1} z,w,v),~~~~~
\lambda\in \mathbb{C}^\ast.
\end{aligned}
\end{align}
Note that from now on our convention will be 
\begin{align}
v= x^4 + ix^5,   ~~~~~~~      w=x^7 + ix^8. 
\end{align}

To be precise, take a conifold 
\begin{align}
uz-wv=0
\end{align}
for example. 
By replacing the conifold tip with a $\mathbb{P}^1$, 
its $A_1$ singularity can be treated as if there is a two-center Taub-NUT space. 
Upon the well-known Taub-NUT/NS5 duality, T-dualizing along the Kaluza-Klein circle $x^6$ makes 
the geometry change to two perpendicular NS5-branes shown in Table \ref{TA}.  
In addition, a complex separation $\Delta x^6 + i\Delta x^9$ arises due to the size of $\mathbb{P}^1$. 
The vanishing two-cycle assumption enables us to set $\Delta x^9=0$.  

\begin{table}[ht]
\begin{center}
\begin{tabular}{|c||c|c|c|c|c|c|c|}
\hline
 &   0123 & 4 & 5 & 6 & 7 & 8&   9 \\ \hline
NS5 & $\circ$   & $\circ$ & $\circ$ &  & & & \\ \hline
NS5' & $\circ$  &  &  & & $\circ$  &  $\circ$ &  \\ \hline
D4    & $\circ$  &  &  & $\circ$ & &   &        \\ \hline
\end{tabular}
\caption[smallcaption]{The NS5/D4-brane configuration for fractional D3-branes 
wrapping the vanishing two-cycle of a conifold after T-duality}
\label{TA}
\end{center}
\end{table}

As far as $X$ concerned, 
near each critical point where $W'(v_i)=0$, the geometry locally looks like 
a conifold. 
With a $\mathbb{P}^1$ resolution on each singularity, after T-duality, 
two NS5-branes having common $0123$ directions are represented as $w=\pm W'(v)$ on $(v,w)$-plane and 
separated along $x^6$ by $l$. 
Furthermore, since 
D5-branes wrapping vanishing two-cycles now become 
D4-branes extending along $01236$, 
the effective four-dimensional gauge coupling reads 
\begin{align}
\begin{aligned}
\frac{1}{g^2_{YM}}=\frac{l}{ 8\pi^2 g_s \ell_s}=
\frac{1}{4\pi g_s}\int_{\mathbb{P}^1} B_{NS}. 
\label{BL}
\end{aligned}
\end{align}
The second equality reveals how 
the K\"ahler moduli of 
$\mathbb{P}^1$ is related to $\Delta x^6$ separation.

\subsection{M-theory lift }

To study the corresponding IR physics, Witten suggested that one should take both large $l$ and $R_{10}=g_s \ell_s$ limit in \eqref{BL} with $\frac{1}{g^2_{YM}}$ being kept finite. This means that the M-cycle opens up and 
Type IIA branes are unified by one smooth M5-brane \cite{MW}. Besides, four-dimensional gauge theory will now be characterized by long-distance informations on the M5-brane. 

\begin{figure}[t]
\begin{center}
\includegraphics[scale=0.5]{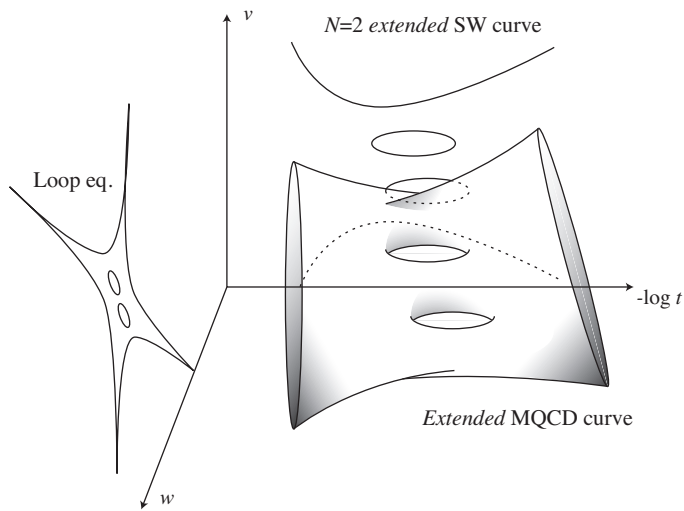}
\end{center}
\caption{The {\it extended } $\Ncal=1$ MQCD curve in $(v,w,t)$-space. 
The projection onto $(v,t)$- and $(v,w)$-plane represents the degenerated {\it extended } SW 
curve and planar loop equation, respectively.}
\label{2}
\end{figure}

In the case without $t_k$ perturbation, 
except for $0123$, the M5-brane wraps a complex curve $\Sigma$, 
holomorphically embedded in $\Mcal_6$ 
($x^{4,5,6,8,9}$ plus the M-cycle $x^{10}$) and parameterized by 
$(w(v), t(v))$ with $t=e^{-s}=\exp -R^{-1}_{10}(x^6 + ix^{10})$.  
$\Sigma$ becomes either a 
Seiberg-Witten curve on 
$(v, t)$-plane or a planar loop equation on $(v, w)$-plane, see Figure \ref{2}. 
More precisely, 
a hyperelliptic curve  
\begin{align}
\begin{aligned}
w^2 - W'(v) w + f_{n-1}(v)=0,
\label{spec}
\end{aligned}
\end{align}
of genus $g=n-1$ on $(v, w)$-plane, which approaches asymptotically to
$ w= W'(v)$ and $w=0$ at $|v|\to \infty$, 
stands for the underlying planar loop equation of $\Ncal=1$ Dijkgraaf-Vafa 
matrix model \cite{Dijkgraaf:2002fc}.

On the other hand, a degenerated Seiberg-Witten curve $t^2 + P_N (v) t +\Lambda^{2N}=0$ 
($\Lambda$: dynamical scale), which 
implies that $N-(g+1)$ mutually local massless monopoles appear, 
is seen on 
$(v, t)$-plane. 
That is, the discriminant now factorizes into 
\begin{align}
\begin{aligned}
\triangle_{SW}& =P_N (v)^2  - 4\Lambda^{2N}= H_{N-n}(v)^2 F_{2n}(v),\\
P_N (v)&=\langle \det (v-\Phi) \rangle ,
\label{SW}
\end{aligned}
\end{align}
where $H_{N-n}$ and $F_{2n}$ are 
polynomials with simple zeros of
degrees $N-n$ and $2n$, respectively. 
It is found \cite{CV} that the extremized 
M-theory curve gives rise to a relation   
\begin{align}
\begin{aligned}
P_N (v)^2  - 4\Lambda^{2N}= \big( W'(v)^2 -f_{n-1}(v) \big)H_{N-n}(v)^2 
\label{SL}
\end{aligned}
\end{align}
between 
\eqref{SW} and $\Ncal=1$ planar loop equation under the constraint  
\begin{align}
P_N (v) \to \prod_{i=1}^{n} (v-v_i)^{N_i}, ~~~~~
\sum_{i=1}^{n} N_i=N, ~~~~~  
\text{as} ~~~~ \Lambda\to 0.
\end{align}
The uniqueness of $P_N(v)$ in \eqref{SL} determines coefficients of the polynomial 
$f_{n-1} (v)$ 
such that all 
glueball vevs in turn get fixed. 
In fact, there is a parallel in the presence of $t_k$ in \eqref{F}. 
As argued 
in \cite {Vafa 2008}, 
parameters $t_k $ and $a_k $, concerning the shape of $\Sigma$, 
are not independent but related to each other at IR. 
Similarly, this is because an on-shell M5-brane 
has to have its volume minimized (minimization of the glueball superpotential).

\section{SUSY/non-SUSY duality}

Let us now turn on arbitrary $t_k$ inside $B$-field such that each group of D4-branes in between NS5-branes will 
no longer have equal length. 
Their lengths vary according to 
\begin{align}
\begin{aligned}
{l(v)} =
2\pi \ell_s \int_{\mathbb{P}^1} B_{NS} (v).
\end{aligned}
\end{align}
If the tree-level superpotential $W(v)$ is dropped out, 
one is left with the so-called $\Ncal=2$ 
$extended$ Seiberg-Witten theory \cite{Ne} whose UV Lagrangian is  
\begin{align}
\begin{aligned}
{\cal{L}}_{UV}=
\frac{1}{2\pi}
\text{Im} \Tr \Big[ \int d^4 \theta~ 
\Fcal'_{UV}(\Phi)
 e^V \bar{\Phi}
+ \int d^2 \theta ~\frac{1}{2} \Fcal_{UV}'' (\Phi)\Wcal_\alpha \Wcal^\alpha \Big],
\end{aligned}
\end{align}
where $V$ is the $\Ncal=1$ vector superfield and $\Fcal_{UV} (\Phi)$ as in \eqref{F} contains higher Casimir terms.

\begin{figure}[t]
\begin{center}
\includegraphics[scale=1]{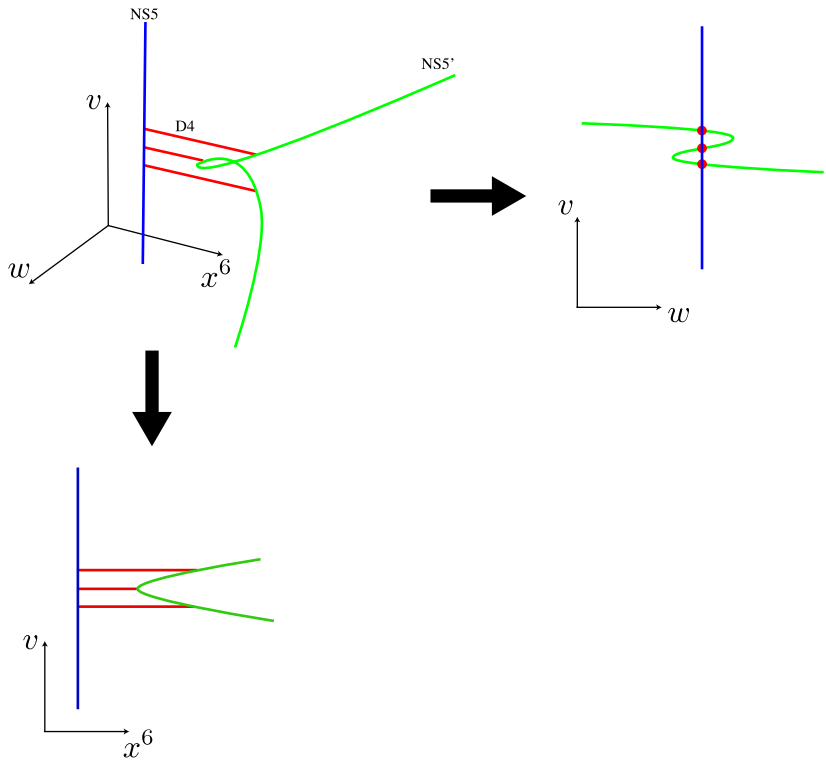}
\end{center}
\caption{\small 
A classical SUSY vacuum in terms of the Type IIA brane picture. Two NS5-branes 
have D4-branes distributed at  
critical loci. Here, $01239$ directions are suppressed. 
The down arrow indicates that a varying $B$-field results in 
differently-sized D4-branes and provides an UV setup for the $extended$ Seiberg-Witten theory. The right arrow implies 
that, despite $t_k$ deformation, 
the underlying CY geometry is still encoded rightly on $(v,x^6)$-plane.}   
\label{2A}
\end{figure}

\begin{figure}[t]
\begin{center}
\includegraphics[scale=0.8]{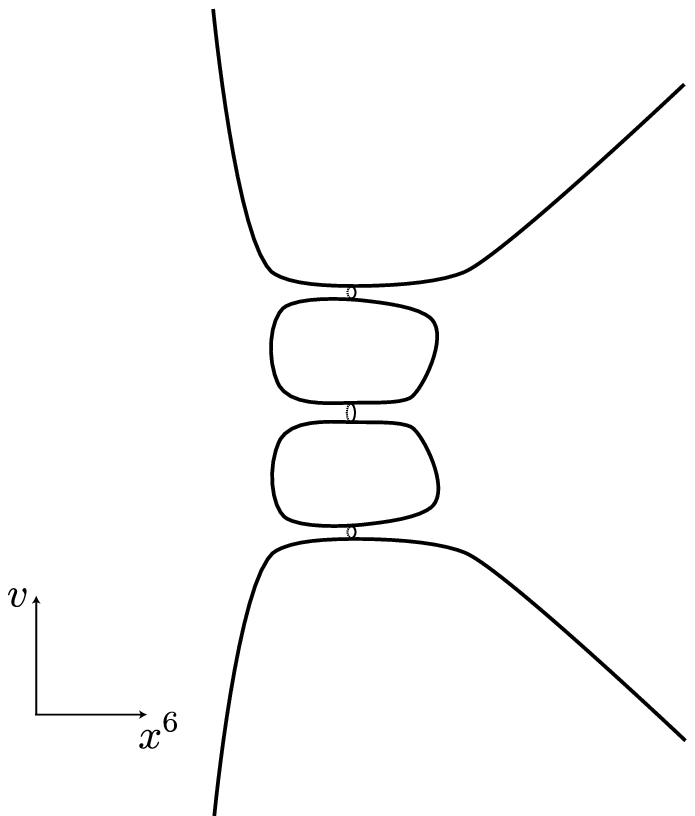}
\end{center}
\caption{The degenerated $extended$ SW curve on $(v,x^6)$-plane. The shape of RHS 
NS5-brane is asymptotically $s\sim v^2 + N\log v$ at large $v$ because of 
the quartic prepotential in \eqref{FS}.}
\label{2R}
\end{figure}

To be explicit, 
an example with the following prepotential and superpotential 
\begin{align}
\begin{aligned}
\Fcal_{UV} (\Phi) &= \Tr\big(\frac{t_2}{12}\Phi^4+\frac{t_1}{6}\Phi^3+\frac{t_0}{2}\Phi^2\big),\\
W(\Phi) &= \Tr\big(a_4\Phi^4+a_3\Phi^3+a_2\Phi^2+a_1\Phi\big),
\label{FS}
\end{aligned}
\end{align}
is plotted in Figure \ref{2A}. 
In spite of $t_k$, the singular CY geometry can still be read off from 
$(v,w)$-plane projection, i.e. $w\big(w-W'(v) \big)=0$.     
However, D4-branes 
are no longer equally-spaced on $(v,x^6)$-plane 
but stretch over the interval 
\begin{align}
\Delta x_6=l(v_i)\propto \Fcal''_{UV} (v_i) 
\label{int}
\end{align}
for $i$-th gauge factor.
For usual $\Ncal=2$ $SU(N)$ SW theory, which is asymptotically free, 
the inverse gauge coupling has a logarithmic one-loop correction. 
This fact is reflected on the bending of 
the MQCD curve, i.e. $t\sim v^N$ for large $v$ and $t$. 
In our case \eqref{FS}, asymptotically we expect that the bending includes an 
extra quadratic term $v^2$, see Figure \ref{2R}.

\begin{figure}[t]
\begin{center}
\begin{tabular}{cc}
\includegraphics{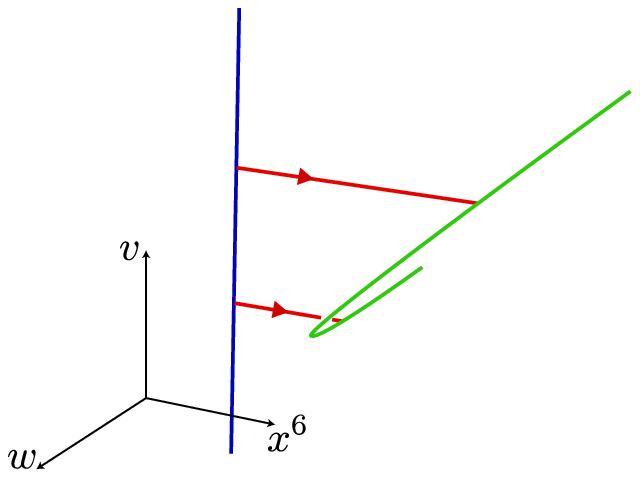}&
\includegraphics{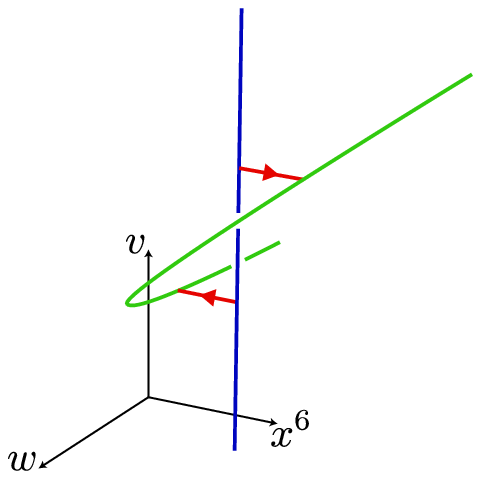}\\
(a) & (b)
\end{tabular}
\end{center}
\caption{Classical Type IIA configurations for (a) SUSY and (b) non-SUSY phases. 
In the non-SUSY case, the orientation of D4-branes is flipped. (a) and (b) differ only 
by a displacement along $x^6$. }
\label{ex}
\end{figure}

Our classical Type IIA 
brane configuration is new in the sense that  
not only $(v,w)$ but $(v,x^6)$ projection yields relatively curved NS5-branes. 
Besides, using these brane setups, one can easily judge 
whether it preserves SUSY or not from 
the intersection of NS5- and D4-branes on $(v,x^6)$-plane. 
This is illustrated in Figure \ref{ex}.

We have assumed a linear $\Fcal_{UV}''$ and a quadratic $W'$ just 
as in \cite{Vafa 2008}.
The authors there showed that how SUSY and non-SUSY vacua occur 
according to $\Fcal_{UV}$. 
The difference between Figure \ref{ex}  
(a) and (b) lies on a displacement along $x^6$, namely, 
the value of $t_0$ in \eqref{F}. Naively, the lower stack of D4-branes in 
Figure \ref{ex} (b) acquires a negative bare gauge coupling for $\Delta x^6 \propto -\frac{1}{g^2_{YM}}<0$ as argued in 
\cite{Vafa 2008}. Rather, this can be interpreted as  
the presence of anti-branes or, in Type IIB language, 
the flip of orientations of $\mathbb{P}^1$'s. 
When lifted to M-theory such that \cite{MW}
\begin{align}
s(v)=\Delta x^6 + i \Delta x^{10} \propto \frac{4\pi }{g^2_{YM}}+ i\frac{\theta}{2\pi}, 
\end{align}
the above fact then emphasizes that 
the M5-brane can no more stay supersymmetric due to its non-holomorphic way of 
embedding with both $s$ and $\bar{s}$. 
As far as the matrix model spectral curve \eqref{spec} concerned, 
anti-eigenvalues (holes) dwelling in $W'(v)=0$ \cite{Dijkgraaf:2002fc} can be thought of as  
the appearance of anti-branes in $(v,x^6)$-space, which do not disturb what happen 
in $(v,w)$-space.

By doing so, spontaneously SUSY breaking vacua can be 
explicitly constructed by means of Type IIA brane configurations like Figure \ref{ex}. 
The terminology ``SUSY/non-SUSY duality" bears similarity to Seiberg duality because they amount to 
crossing NS5-branes and thereby changing the coupling constant.

\subsection{$\Ncal =1$ effective superpotential}

Now, let us see how the effective superpotential gets 
modified in the presence of $t_k$.  
For $\Ncal=1$ gauge theory, the M-theory approach to 
deriving the effective superpotential is initiated by Witten \cite{Witten}. 
He suggested the following integral
\begin{align}
W_{ MQCD}=\int_B \Omega_3,
\label{mqcd}
\end{align}
where $\Omega_3\equiv dv \wedge
dw \wedge \frac{dt}{t}$ is a holomorphic three-form. $B$ is a
three-manifold having two boundaries, i.e. the previously-defined $\Sigma$ 
and a reference surface $\Sigma_0$ homologous to 
$\Sigma$.

\begin{figure}[t]
\begin{center}
\includegraphics[scale=0.6]{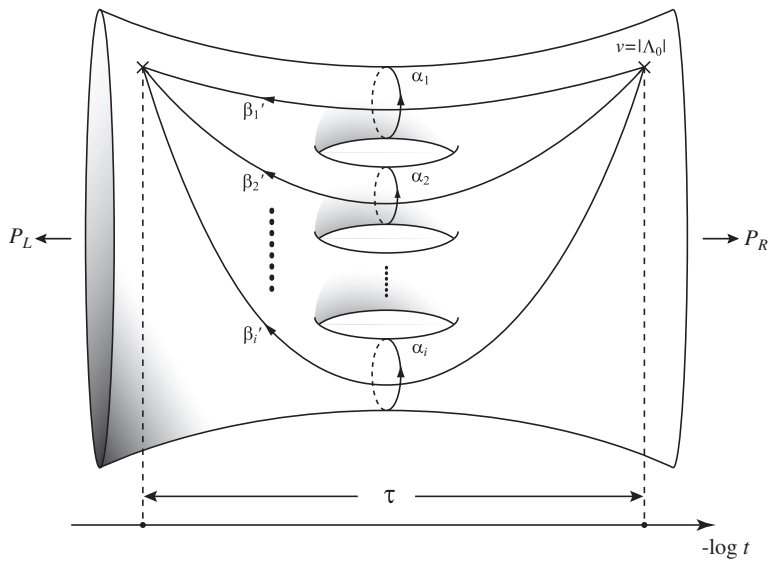}
\end{center}
\caption{Cycles for integrals in the glueball 
superpotential. $\beta'_i$-cycles are regulated at $|v|=\Lambda_0$.}
\label{3}
\end{figure}

If there exists a two-form $\Omega_2$ which satisfies
\begin{align}
\Omega_3=d\Omega_2,
\end{align}
then \eqref{mqcd} can be written as
\begin{align}
W_{ MQCD}=W(\Sigma)-W(\Sigma_0),
\end{align}
where $W(\Sigma)=\int_{\Sigma}\Omega_2$ and  $W(\Sigma_0)=\int_{\Sigma_0}\Omega_2$.  
Since $W(\Sigma_0)$ is physically irrelevant, the 
effective superpotential reduces to 
\begin{align}
W_{ eff} 
 = \int_{\Sigma}\Omega_2, ~~~~~~~\Omega_2=-wdv \wedge \frac{dt}{t}.
\label{Weff}
\end{align}
Now, it is straightforward that 
\begin{align}
\begin{aligned}
W_{ eff} =-\int_{\Sigma} wdv \wedge \frac{dt}{t}
=\sum_{i} \left(
\oint_{\alpha_i}\frac{dt}{t} \int_{\beta_i}wdv
-\int_{\beta_i}\frac{dt}{t} \oint_{\alpha_i}wdv \right)
\label{e}
\end{aligned}
\end{align}
upon making use of Riemann's 
bilinear identity. 
Here, $\alpha_i$'s denote cycles around cuts while 
$\beta_i$'s are paths connecting $P_R$ and $P_L$, see Figure \ref{3}.

Because $t\sim v^{N_i}$ near the neighborhood of each cut 
($N_i$: number of D4-branes attached on the cut), one has 
\begin{align}
\oint_{\alpha_i} \frac{dt}{t} = N_i.
\label{t}
\end{align}
Next, since integrals over $\beta_i$ naively diverge, 
it is necessary to introduce a 
cut-off scale at $|v|=\Lambda_0$ for regularization. 
The integral $\int _{\beta'}\frac{dt}{t}$ is nothing but a line integral
over the coordinate $s$, that is, 
it just gives a regularized complex 
separation 
between two NS5-branes or 
the gauge coupling on the compactified
D4-brane. Therefore, the bare Yang-Mills coupling constant
$\alpha_i (\Lambda_0)=
\frac{\theta}{2\pi}+ \frac{4\pi i}{g^2_{YM}}$ 
evaluated at $\Lambda_0$ is related to $\tau_i=\int_{\beta_i'}\frac{dt}{t}$ 
by 
\begin{align}
\begin{aligned}
\frac{\tau_i}{2\pi i R_{10}}=-\alpha_i, ~~~~~~~
R_{10}=g_s \ell_s.
\end{aligned}
\end{align}

Plugging these into \eqref{e}, we obtain the 
effective superpotential ($R_{10}=1$)
\begin{align}
W_{eff}
= \sum_{i} \left(
    N_i\Pi_i + 2\pi i \alpha_i S_i
    \right),
\label{Weff of Si}
\end{align}
where the glueball $ S_i \equiv \oint_{\alpha_i}wdv$ and 
$\frac{\partial \Fcal_0}{\partial S_i}=
\Pi_i \equiv
\int_{\beta_i'}wdv$ stand for dual periods in the context of special geometry. 
With $t_k$ perturbation, from \eqref{int} we find that 
\eqref{e} can be   
immediately generalized into 
\begin{align}
\begin{aligned}
W_{eff}=\sum_i \big( N_i \Pi_i 
+ 2\pi i\oint_{\alpha_i} \alpha_i(v)w(v)dv \big).
\label{sup}
\end{aligned}
\end{align}
This 
reproduces precisely what derived by Aganagic et al. in \cite{Vafa 2008}.

\subsection{Partial SUSY breaking from $\Ncal=2$ to $\Ncal=1$}

\begin{figure}[t]
\begin{center}
\begin{tabular}{ccccc}
\includegraphics[scale=0.5]{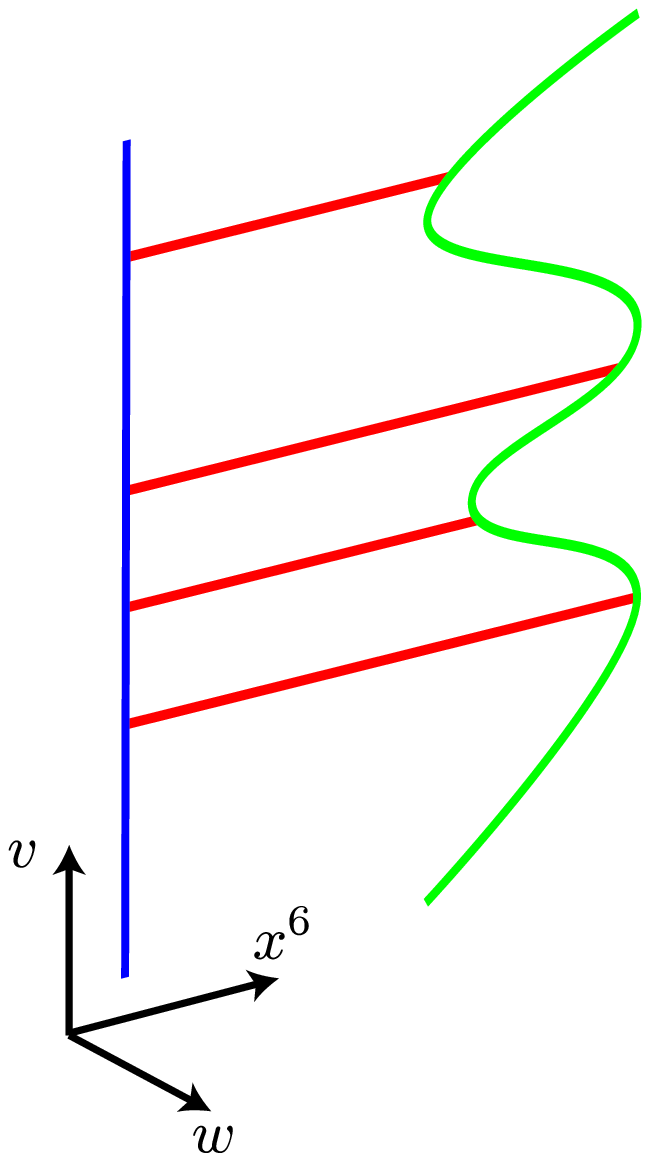}&\hspace*{1cm}&
\includegraphics[scale=0.4]{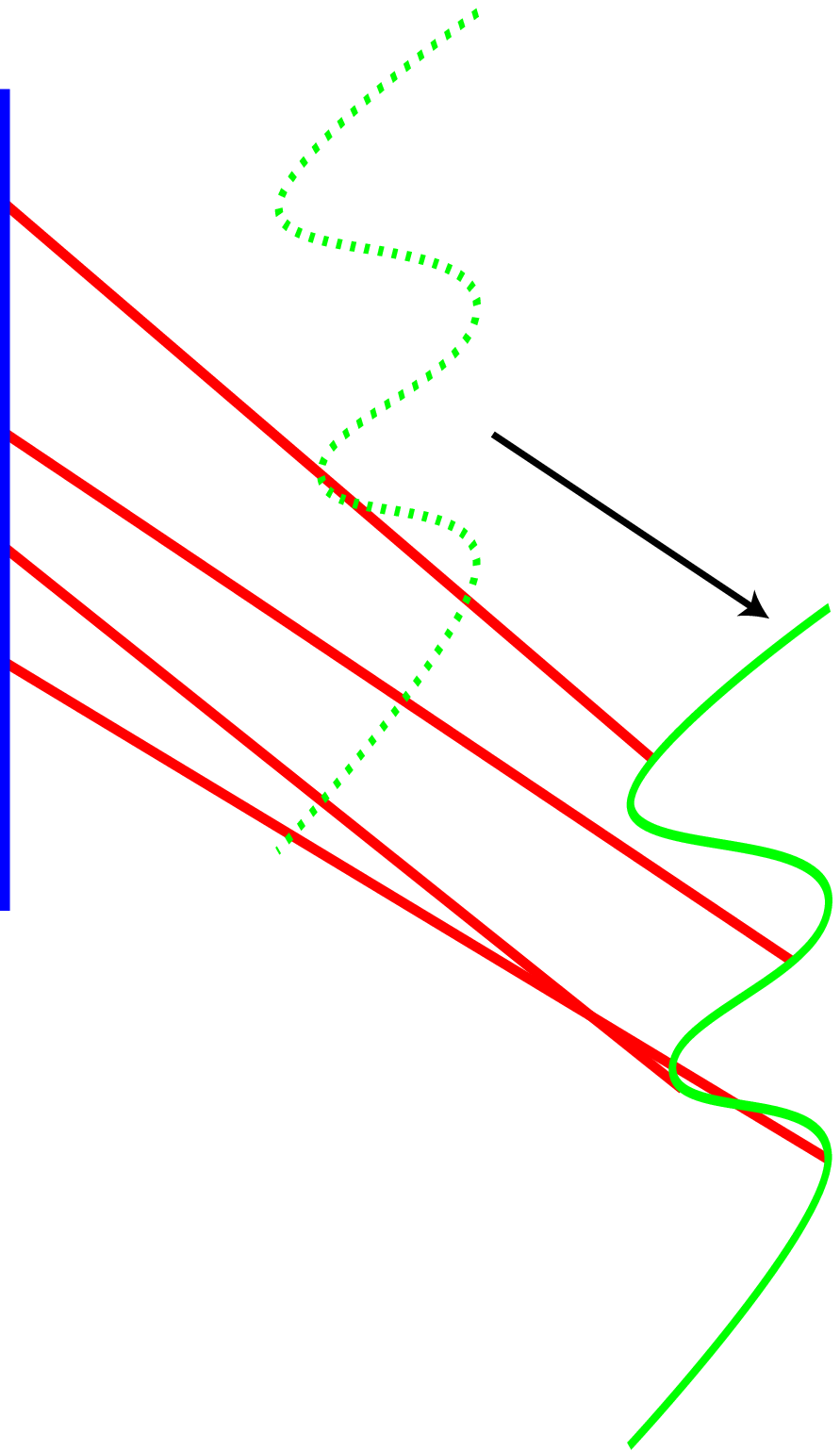}&\hspace*{1cm}&
\includegraphics[scale=0.4]{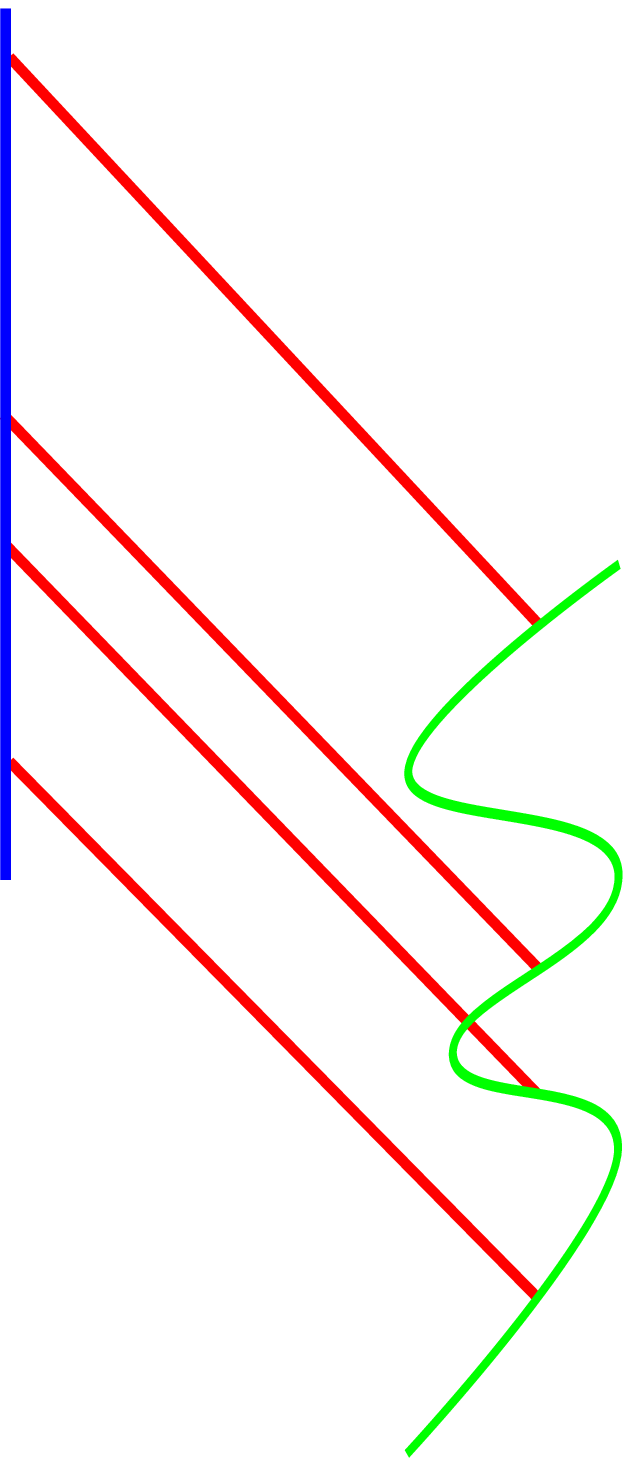}\\
(a)&&(b)&&(c)
\end{tabular}
\end{center}
\caption{Partial SUSY breaking from $\Ncal=2$ configuration to $\Ncal=1$ one. 
Turning on FI parameters, SUSY of 
the $extended$ $\Ncal=2$ theory gets completely broken (off-shell) temporarily. 
SUSY is recovered (on-shell) again at critical loci where $\Fcal''(\Phi)=W'(\Phi)=0$, but 
only $\Ncal=1$ is now preserved. }
\label{Partial SUSY}
\end{figure}

Finally, we comment on the partial SUSY breaking from 
$\Ncal=2$ to $\Ncal=1$ through the above 
brane picture. The partial SUSY breaking is discussed in \cite{Antoniadis:1995vb,Ferrara:1995xi,Partouche:1996yp}
for Abelian gauge group. The non-Abelian generalization is 
well investigated and established 
by authors of \cite{Fujiwara:2004kc}. 
According to these early works, $\Ncal=2$ theory is firstly perturbed by introducing a 
general prepotential of the form \eqref{F} and 
its SUSY is broken down to $\Ncal=1$ thereof upon adding 
Fayet-Iliopoulos (FI) parameters.

As mentioned throughout this paper, the prepotential $\Fcal_{UV}(\Phi)$ 
describes how NS5-branes get deformed in $(v,x^6)$-space, 
see Figure \ref{Partial SUSY} (a). 
Three FI parameters correspond to the relative position of 
two NS5-branes 
in $(x^7,x^8,x^9)$-space. 
Henceforth, turning on FI parameters means that two NS5-branes are separated from 
each other in 
$(x^7,x^8,x^9)$-space.
In addition, a new direction by which the bare coupling constant is measured should be 
defined due to the presence of FI parameters. 
In Figure \ref{Partial SUSY} (a), initial positions of D4-branes are  
not fixed. But if D4-branes still remain at their initial positions, they 
become non-parallel when the curved NS5-brane is pulled along $w$, see Figure \ref{Partial SUSY} (b). 
In other words, SUSY can no longer 
be maintained for an off-shell choice of $\Phi$ vev.  
To recover SUSY, D4-branes should be re-distributed appropriately at critical loci, i.e. $\Fcal''(\Phi)=0$ (on-shell condition). Now, since SUSY is recovered to $\Ncal=1$, 
we can as well recognize the tree-level 
superpotential as $W(\Phi)=\Fcal'(\Phi)$ 
with coefficients rescaled suitably, see Figure \ref{Partial SUSY} (c). 
To this end, 
the partial SUSY breaking mechanism can thus be understood pictorially from the 
{\it extended} brane configuration.

The above argument is valid only for the classical (UV) theory. 
In order to extend this picture to the full quantum theory, 
we need to replace the NS5/D4 system with a MQCD curve. 
Pulling out one NS5-brane then corresponds to
deforming the curve. 
The projective information of the MQCD curve 
contains both the {\it extended} SW curve and 
loop equation (or, alternatively, 
generalized Konishi anomaly equation) of $\Ncal=1$ theory. 
Therefore, it is interesting to see how these aspects transform 
according to the partial SUSY breaking deformation of the curve.

\section{Conclusion and discussion}
So far, we have shown how the SUSY/non-SUSY duality proposed 
by Aganagic et al. can have a corresponding Type IIA brane picture. 
Apart from conventional ones, 
where anti-branes are wrapped on a CY by hand, the setup here involves 
changing the orientation of local two-cycles through a varying 
background NS-flux. This dose work because $B$-field gives a 
K\"ahler moduli $\Delta t\sim B_{NS}$ of arbitrary sign 
to shrinking 
two-cycles and hence 
controls their flops. 
On Type IIA side, we interpret this background as two crossing NS5-branes 
where $\overline{\text{D4}}$-branes appear naturally for flipped orientations.  
Consequently, simultaneous presence of D4- and $\overline{\text{D4}}$-branes 
soon suggests a way to realize 
various kinds of SUSY/non-SUSY vacua via adjusting parameters the NS-flux contains. 
Moreover, curved NS5-branes on $(v,t)$-plane with $\Ncal=2$ SUSY 
correspond to what 
has been known as the $extended$ Seiberg-Witten theory. 
One can further add FI parameters to partially 
break $\Ncal=2$ down to $\Ncal=1$. 
Resorting to Type IIA brane pictures, we see this process is 
clearly visualized in Figure \ref{Partial SUSY}. The final $\Ncal=1$ vacuum 
is arrived at once the tree-level superpotential $W(\Phi)$ takes the form of $\Fcal'(\Phi)$.

We also considered M-theory lift. Without $t_k$ perturbation, 
the M-theory curve itself is either a degenerated Seiberg-Witten curve on $(v,t)$-plane or 
a loop equation of DV matrix model on $(v,w)$-plane. 
Though adding higher $t_k$ terms has no effect on the planar loop equation, 
we find that $\Ncal=1$ effective superpotential which involves $\beta$-cycles on 
$(v,t)$-plane gets modified. 
In particular, 
it seems that 
the above partial SUSY breaking process can be described by deforming 
one given M-theory curve in order to incorporate quantum effects. 
It is thus of interest to 
compare this observation with field theory results found in \cite{Itoyama:2007rr}. We leave these problems to future works.

\subsection*{Acknowledgements}
We are grateful to K.~Hashimoto for helpful discussions. 
KO would like to thank T.~Higaki and K.~Maruyoshi for useful comments.
KO is supported in part by Grant-in-Aid for Scientific 
Research (No.19740120) from the Ministry 
of Education, Culture, Sports, Science and Technology. 
TST is supported in part by the postdoctoral program at RIKEN. 
 



\begin{thebibliography}{10}

\bibitem{Vafa 2008}
M. Aganagic, C. Beem, J. Seo and C. Vafa, 
{\it ``Extended Supersymmetric Moduli Space and a SUSY/Non-SUSY Duality,''} 
arXiv:0804.2489 [hep-th] 


\bibitem{Vafa 2006}

M. Aganagic, C. Beem, J. Seo and C. Vafa, 
{\it ``Geometrically induced 
metastability and holography,"} 
Nucl. Phys. B {\bf  789} 382 (2008) [arXiv:hep-th/0610249] 
\bibitem{Mar}

J. Marsano, K. Papadodimas and M. Shigemori, 
Nucl. Phys. B {\bf 789} 294 (2008)  
arXiv:0705.0983 [hep-th]; arXiv:0801.2154 [hep-th] 
\bibitem{Vafa 2007}

M. Aganagic, C. Beem and B. Freivogel, 
{\it ``Geometric Metastability, Quivers 
and Holography,"} Nucl. Phys. {\bf B 795} 291 (2008) 
arXiv:0708.0596 [hep-th] 

\bibitem{Hollands:2008cs}
  L.~Hollands, J.~Marsano, K.~Papadodimas and M.~Shigemori, 
{\it ``Nonsupersymmetric Flux Vacua and Perturbed N=2 Systems,"}  
arXiv:0804.4006 [hep-th]

\bibitem{Ooguri:2006bg}
  H.~Ooguri and Y.~Ookouchi, 
{\it ``Meta-stable supersymmetry breaking vacua on intersecting branes,''} 
Phys.\ Lett.\  B {\bf 641}, 323 (2006) [arXiv:hep-th/0607183] 




\bibitem{me1}
S. Franco, I. Garcia-Etxebarria and A. M. Uranga, {\it ``Non-supersymmetric meta-stable
vacua from brane configurations,"} JHEP {\bf 0701} 085 (2007) [arXiv:hep-th/0607218] 

\bibitem{me2}
I. Bena, E. Gorbatov, S. Hellerman, N. Seiberg and D. Shih, {\it ``A note on (meta)stable
brane configurations in MQCD,"} JHEP {\bf 0611} 088 (2006) [arXiv:hep-th/0608157] 




\bibitem{Intriligator:2006dd}
  K.~Intriligator, N.~Seiberg and D.~Shih,
  JHEP {\bf 0604}, 021 (2006)
  [arXiv:hep-th/0602239] 




\bibitem{GK}
A. Giveon and D. Kutasov, 
{\it ``Brane dynamics and gauge theory,"}
Rev. Mod. Phys. {\bf 71} 983 (1999)  
[arXiv:hep-th/9802067] 
\bibitem{HW}
A. Hanany and E. Witten, {\it ``Type IIB superstrings, BPS monopoles, and
three-dimensional gauge dynamics,"} Nucl. Phys. B {\bf 492} 152 (1997)
[arXiv:hep-th/9611230] 

\bibitem{a}
M. Bershadsky, C. Vafa and V. Sadov, 
{\it ``D strings on D manifolds,"} 
Nucl. Phys. B {\bf 463} 398 (1996) 
[arXiv:hep-th/9510225] 


\bibitem{a1}
H. Ooguri and C. Vafa, 
{\it ``Two-dimensional black hole and singularities of CY manifolds,"} 
Nucl. Phys. B {\bf 463} 55 (1996) 
[arXiv:hep-th/9511164] 


\bibitem{a2}
A. Karch, D. Lust and D. J. Smith, 
{\it ``Equivalence of geometric engineering and Hanany-Witten via fractional branes,"}
Nucl. Phys. B {\bf 533} 348 (1998)  
[arXiv:hep-th/9803232] 

\bibitem{a3}

K. Dasgupta, K. Oh and R. Tatar, 
{\it ``Geometric transition, large N dualities and MQCD dynamics,"} 
Nucl. Phys. B {\bf 610} 331 (2001)  
[arXiv:hep-th/0105066] 


\bibitem{MW}

E. Witten, 
{\it ``Solutions of four-dimensional field theories via M theory,"}
Nucl. Phys. B {\bf 500} 3 (1997) 
[arXiv:hep-th/9703166] 


\bibitem{Dijkgraaf:2002fc}
  R.~Dijkgraaf and C.~Vafa,
  Nucl.\ Phys.\ B {\bf 644} 3 (2002)
  [arXiv:hep-th/0206255];
  Nucl.\ Phys.\ B {\bf 644} 21 (2002)
  [arXiv:hep-th/0207106];
  [arXiv:hep-th/0208048]

\bibitem{CV}
F. Cachazo, K. Intriligator, and C. Vafa, 
{\it ``A Large N Duality via a Geometric Transition,"} 
Nucl. Phys. B {\bf 603} 3 (2001) [arXiv:hep-th/0103067] 



\bibitem{Ne}
A. Marshakov and N. Nekrasov, 
{\it ``Extended Seiberg-Witten theory and
integrable hierarchy,"} JHEP {\bf 0701} 104 (2007) [arXiv:hep-th/0612019] 


\bibitem{Witten}
E. Witten, 
{\it ``Branes and the dynamics of QCD,"} 
Nucl. Phys. B {\bf 507} 658 (1997) 
[arXiv:hep-th/9706109] 


\bibitem{Antoniadis:1995vb}
  I.~Antoniadis, H.~Partouche and T.~R.~Taylor,
  Phys.\ Lett.\  B {\bf 372}, 83 (1996)
  [arXiv:hep-th/9512006];
  I.~Antoniadis and T.~R.~Taylor,
  Fortsch.\ Phys.\  {\bf 44}, 487 (1996)
  [arXiv:hep-th/9604062]

\bibitem{Ferrara:1995xi}
  S.~Ferrara, L.~Girardello and M.~Porrati,
  Phys.\ Lett.\  B {\bf 376}, 275 (1996)
  [arXiv:hep-th/9512180]

\bibitem{Partouche:1996yp}
  H.~Partouche and B.~Pioline,
  Nucl.\ Phys.\ Proc.\ Suppl.\  {\bf 56B}, 322 (1997)
  [arXiv:hep-th/9702115]
 
\bibitem{Fujiwara:2004kc}
  K.~Fujiwara, H.~Itoyama and M.~Sakaguchi,
  Prog.\ Theor.\ Phys.\  {\bf 113}, 429 (2005)
  [arXiv:hep-th/0409060];
  Nucl.\ Phys.\  B {\bf 723}, 33 (2005)
  [arXiv:hep-th/0503113];
  K.~Fujiwara,
  Nucl.\ Phys.\  B {\bf 770}, 145 (2007)
  [arXiv:hep-th/0609039] 





%
%



  





\bibitem{Itoyama:2007rr}
  H.~Itoyama and K.~Maruyoshi,
  Phys.\ Lett.\  B {\bf 650}, 298 (2007) arXiv:0704.1060 [hep-th];
  Nucl.\ Phys.\  B {\bf 796}, 246 (2008) 
arXiv:0710.4377 [hep-th]


\end{thebibliography}
\end{document}